\documentclass[10pt,onecolumn,letterpaper]{article}

 \usepackage{cvpr}              

\usepackage[dvipsnames]{xcolor}

\definecolor{cvprblue}{rgb}{0.21,0.49,0.74}
\usepackage[pagebackref,breaklinks,colorlinks,citecolor=cvprblue]{hyperref}
\usepackage{multirow}

\def\paperID{*****} 
\def\confName{CVPR}
\def\confYear{2025}

\title{Hyperspectral image reconstruction by deep learning with super-Rayleigh speckles}

\author{Ziyan Chen\\
School of Mathematics and Physics,Hubei Polytechnic University, Huangshi, Hubei, 435000, China\\
{\tt\small chenziyan@hbpu.edu.cn}
\and
Zhentao Liu*
, Jianrong Wu, Shensheng Han\\
Key Laboratory for Quantum Optics of CAS, Shanghai Institute of Optics and Fine Mechanics, \\Chinese Academy of Sciences, Shanghai, 201800, China\\
{\tt\small ztliu@siom.ac.cn}
}

\begin{document}
\maketitle
\abstract{ \rm{
 Ghost imaging via sparsity constraints (GISC) spectral camera modulates the three-dimensional (3D) hyperspectral image into a two-dimensional (2D) compressive image with speckles in a single shot. It obtains a 3D hyperspectral image (HSI) by reconstruction algorithms. The rapid development of deep learning has provided a new method for 3D HSI reconstruction. Moreover, the imaging performance of the GISC spectral camera can be improved by optimizing the speckle modulation. In this paper, we propose an end-to-end GISCnet with super-Rayleigh speckle modulation to improve the imaging quality of the GISC spectral camera. The structure of GISCnet is very simple but effective, and we can easily adjust the network structure parameters to improve the image reconstruction quality. Relative to Rayleigh speckles, our super-Rayleigh speckles modulation exhibits a wealth of detail in reconstructing 3D HSIs. After evaluating 648 3D HSIs, it was found that the average peak signal-to-noise ratio increased from 27 dB to 31 dB. Overall, the proposed GISCnet with super-Rayleigh speckle modulation can effectively improve the imaging quality of the GISC spectral camera by taking advantage of both optimized super-Rayleigh modulation and deep-learning image reconstruction, inspiring joint optimization of light-field modulation and image reconstruction to improve ghost imaging performance.}
}

\section{Introduction} 
\rm {Ghost imaging (GI) is an imaging technique that can obtain image information through the intensity correlation of optical fields between the object path and the reference path \cite{ cheng2004incoherent}. It encodes the image information into the intensity fluctuations of light fields, and then the high-dimensional data can be reconstructed from the low-dimensional detecting measurements\cite{duarte2008single, hardy2013computational, rohtua, Howland:11, han2018review, Aguilar:22, sun2016single, gong2016three, liu2016spectral}.  Early GI obtains two-dimensional (2D) images reconstructed from one-dimensional (1D) signals\cite{cheng2004incoherent, duarte2008single, hardy2013computational}. Lately, GI technology has been applied to obtain three-dimensional (3D) images\cite{rohtua, Howland:11, han2018review, Aguilar:22, sun2016single, gong2016three}, and ghost imaging via sparsity constraints (GISC) spectral camera is a typical case\cite{liu2016spectral}. 

GISC spectral camera modulates the 3D hyperspectral image (HSI) into 2D spatial intensity fluctuations of light fields, which enables capturing the 3D signals information in a single shot. It provides a specific solution for spectral imaging of dynamic processes, which can be widely applied in remote sensing, microscopy, and astronomy. There are mainly two kinds of methods for reconstructing 3D HIS from the 2D detection image: one is correlation algorithms \cite{cheng2004incoherent, ferri2010differential, tong2021preconditioned} and the other is the compressive sensing (CS)  \cite{donoho2006compressed, candes2006robust, eldar2012compressed}algorithms.  Conventional GI correlation algorithms, such as differential GI (DGI) \cite{ferri2010differential}, although has a fast reconstruction speed. Still, it suffers from low reconstruction quality, especially with a low sampling rate and signal-to-noise ratio. Though CS algorithms can get higher reconstruction quality than conventional GI correlation algorithms, it's time-consuming because of the interactive process. The reconstructed results are unsatisfactory when the detection image's signal-to-noise ratio (SNR) is low. In recent years, deep learning (DL) has provided new opportunities and tools for computational imaging\cite{barbastathis2019use}. Many researchers have applied it to ghost imaging and have achieved good performance \cite{wu2020sub, Zhu:20, hu2020denoising, he2018ghost, li2020compressive, wangfei, czydgi}. Most of them deal with the 2D reconstruction in GI \cite{wu2020sub, Zhu:20, hu2020denoising,  he2018ghost, li2020compressive, wangfei}, the reconstruction of 3D ghost imaging based on deep learning is rarely done\cite{czydgi}.  

In the \cite{czydgi}, V-DUnet has successfully improved the reconstruction quality of 3D HSIs in the GISC spectral camera by setting both DGI results and the detected measurements as the network’s input. However, when tested on several hundred hyperspectral images, it was found that their average peak signal-to-noise ratio (PSNR) was about 27 dB, which indicates that there is still room for improvement.  On the other hand, speckle patterns play an essential role in ghost imaging. Much research on optimizing speckle patterns has been developed to improve the sampling efficiency and the reconstruction quality\cite{spe1, spe2, spe3, spe4, speckle1}. Usually, speckle patterns show Rayleigh statistics, and its contrast is theoretically equal to 1. Compared to Rayleigh speckles, super-Rayleigh speckles have a higher contrast\cite{chaoruili1,chaoruili2}. Liu et al. proposed snapshot spectral ghost imaging with super-Rayleigh speckles, demonstrating superior noise immunity and higher imaging quality at low sampling rates compared to universal Rayleigh speckle patterns\cite{chaoruili2}. In this article, we proposed an end-to-end GISCnet with super-Rayleigh speckle modulation to optimize the GISC spectral camera’s imaging quality. 648 3D HSIs have been tested for analyzing imaging ability. Compared with work \cite{czydgi}, the average PSNR of imaging results has been increased by about 4 dB. It verified that the imaging ability of the GISC spectral camera has dramatically improved by super-Rayleigh modulation and GISCnet reconstruction, which will promote its applications in biological imaging and remote sensing.

\section{System of GISC spectral camera}

\begin{figure*}[htb]
\centering
\includegraphics[width=12cm,angle=0]{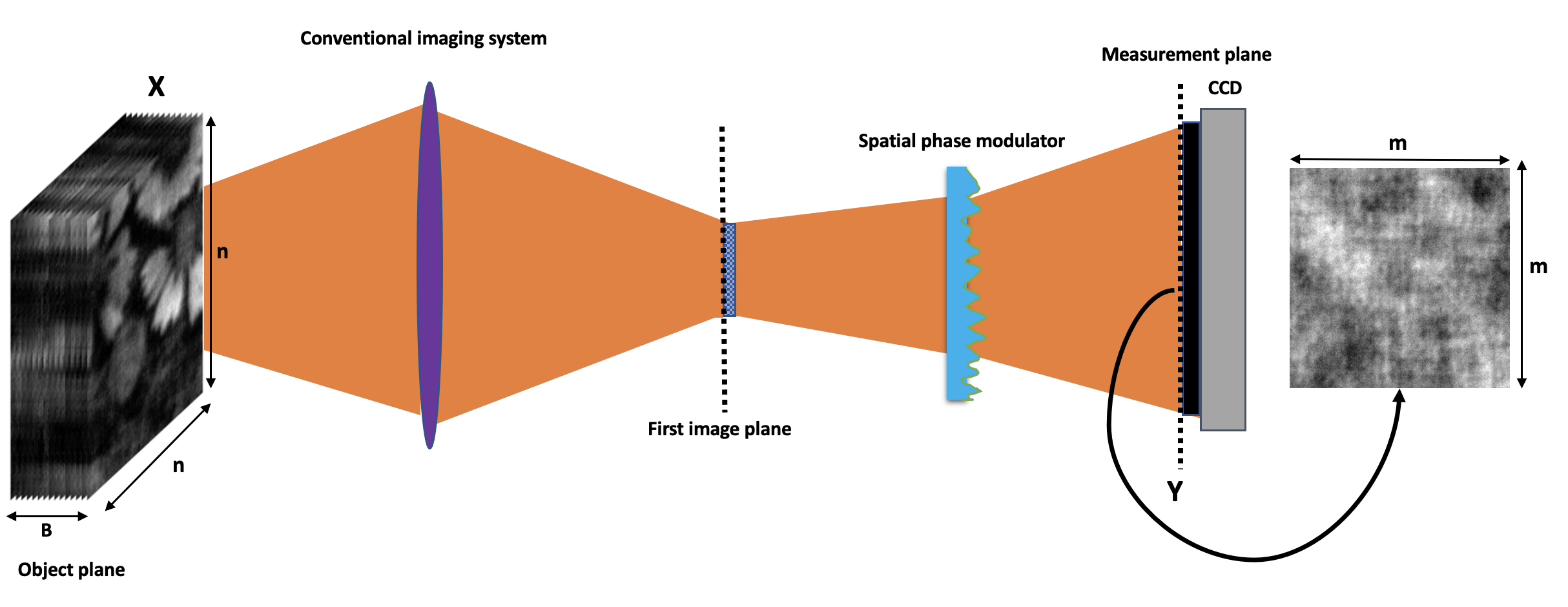}
\caption{The schematic of GISC spectral camera. }
\label{f1}
\end{figure*}
In Fig.\ref{f1}, we can see that the system contains three modules: the front imaging module, the modulation module, and the detection module. The first imaging plane collects lights from the 3D HSI ${X}$ 
by a conventional imaging system (the front imaging module). Then the light fields in the first imaging plane are modulated by a spatial random phase modulator (the modulation module); finally,  a CCD detector is used to record the modulated imaging speckle patterns ${\bf Y}$ (the detection module).   

The calibrated speckle patterns are necessary for reconstitution and should be pre-determined before the imaging process. It can be obtained by scanning along the spatial and spectral dimensions with a monochromatic point source on the object plane. One can refer to \cite{liu2016spectral, czydgi} to better understand the GISC camera's calibration process. The 3D HSI can be reconstructed by calculating the intensity correlation between the calibrated speckle patterns and imaging speckle patterns \cite{liu2016spectral}. Meanwhile, the imaging process can be written into a matrix form as \cite{han2018review}
\begin{eqnarray}
{y} =\Phi {x} + \epsilon,
 \label{eq1}
\end{eqnarray}
in which ${x}  \in \mathbb{R} ^{nnB\times1}$ is reshaped from the HSI  ${\bf X} \in \mathbb{R} ^{n \times n \times B}$ , ${y}  \in \mathbb{R} ^{mm\times1} $ is reshaped from the measurement image ${\bf Y}\in \mathbb{R} ^{m \times m}$. $\epsilon \in \mathbb{R} ^{mm\times1}$ represents the noise of the system. $\Phi \in \mathbb{R} ^{mm\times nnB  }$ is the sensing matrix, and each column vector in $\Phi $ presents a calibrated speckle intensity pattern corresponding to one pixel in HSI. 

\section{The proposed framework}
\begin{figure*}[htb]
\centering
\includegraphics[width=13cm,angle=0]{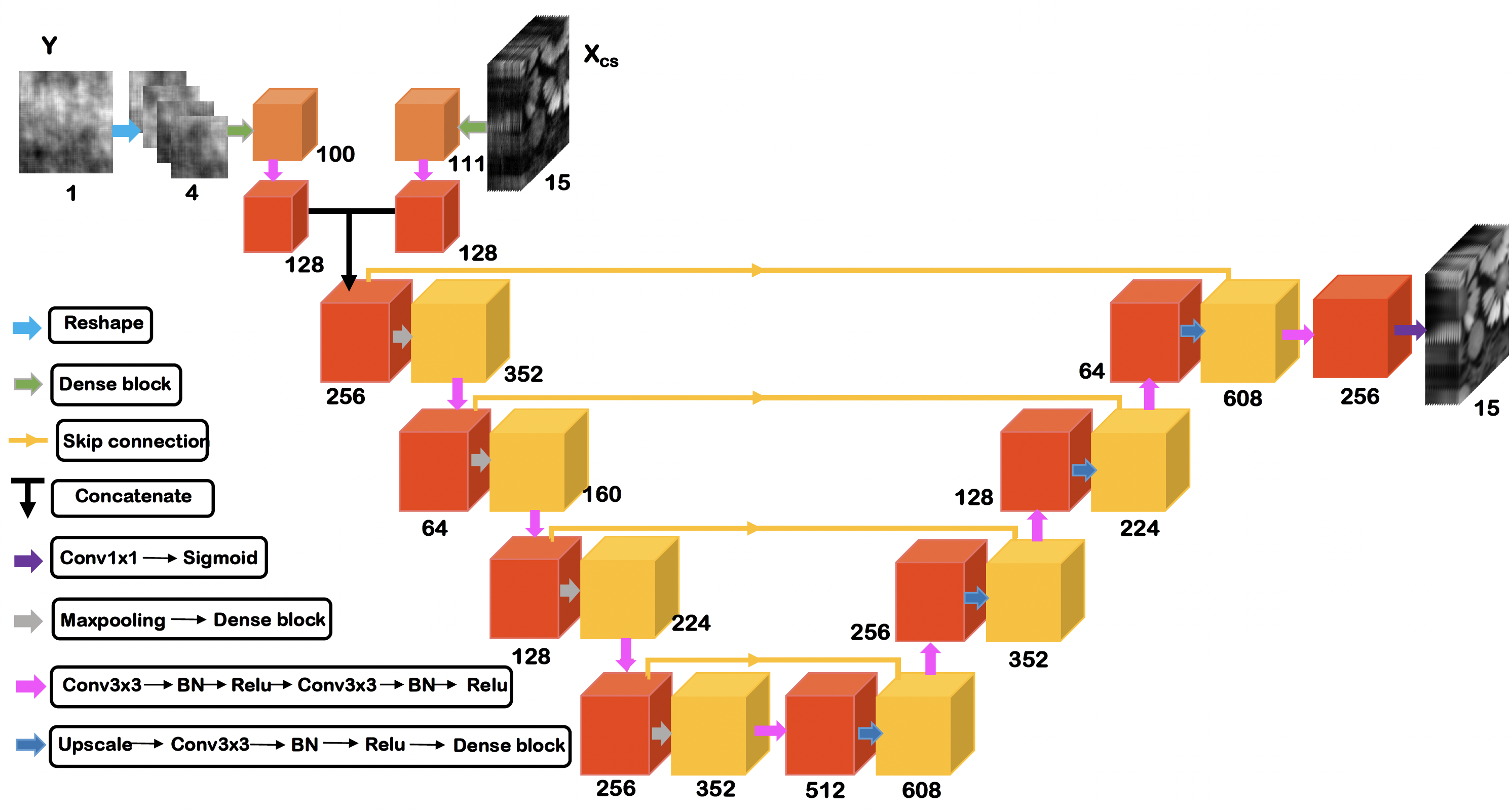}
\caption{The architecture of the proposed GISCnet. BN, batch normalization; Conv $3 \times 3$, convolution with filter size $3 \times 3$; Conv $1 \times 1$, convolution with filter size $1 \times1$; Dropout, dropout rate is $0.05$; Relu, rectified linear unit;  Maxpooling , stride ($2, 2$); Upscale, factor $2$.}
\label{f2}
\end{figure*}
Our proposed framework is presented in Fig.\ref{f2}. It can be seen that the backbone of the GISCnet is UD block. The combination mode of Unet \cite{ronneberger2015u} and dense block \cite{huang2017densely} in UD block is shown in Fig.\ref{f3}. The proposed GISCnet includes two inputs: the measurement image ${\bf Y}$ with $288 \times 288$ pixels recorded by the CCD, and the preprocessed reconstructed CS image $\bf {X_{cs}}$ with size $144 \times 144 \times 15$\cite{PICHCS}. Before feeding $\bf Y$ and $\bf {X_{cs}}$ into the first UD block,  ${\bf Y}$ is reshaped into four patches with size $144 \times 144 \times 4$. Those patches and $\bf {X_{cs}}$ pass through two convolutional blocks respectively and finally concatenate together and feed into the first UD block. The output channel number of the ith UD block is

  \begin{figure}[htpb]
\centering
\includegraphics[width=6cm,angle=0]{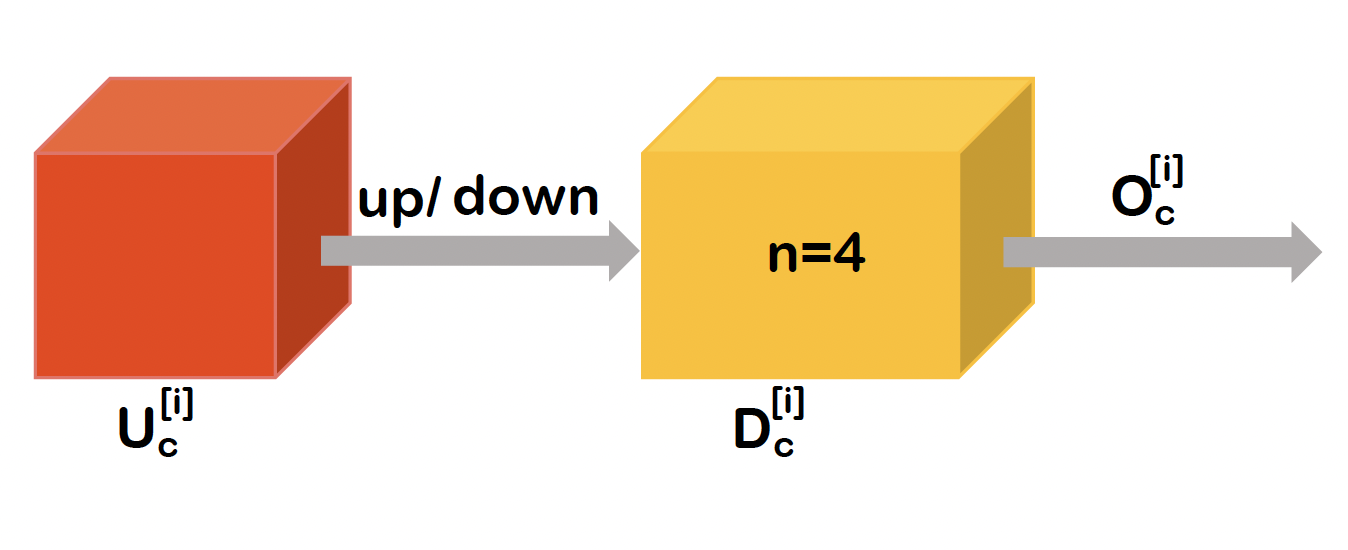}
\caption{The architecture of the ith UD block.}
\label{f3}
\end{figure}  

\begin{eqnarray}
O^{[i]}_{c}=a \times U^{[i]}_{c}+(n-1)\times D_c^{[i]}
  \label{eq_ch3_1}
\end{eqnarray} 
where $D_{c}^{[i]}$ is the growth rate parameter of the dense block,  $U_{c}^{[i]}$ represents the number of Unet channels in the ith UD block, $n$ represents the number of layers of the dense block,  $a$ is a parameter related to the sampling characteristics of the ith Unet layer. If the Unet layer in the current UD block is in the down-sampling stage, $a=1$. And $a=1⁄2$ in the up-sampling stage. Different from the value of $D_{c}^{[i]}$  changing with the location of the Unet layer in V-DUnet, the value of $D_{c}^{[i]}$ is a fixed value in GISCnet. We set $D_ c^{[i]}=32$ and $n=4$. Additionally, FFDNet \cite{zhang2018ffdnet} is applied in the training process to denoise $\bf Y$ and $\bf {X_{cs}}$ in the GISCnet.

The loss function of GISCnet  is     
\begin{eqnarray}
  & Loss=&\alpha \| X-\hat {X}\|_1  
 \label{eq2}
\end{eqnarray}  
here we set $\alpha=50$,  $\| \ \|_1$ represents the L1 norm. $X$ represents the ground truth of the original HSI while $\hat {X}$ is the corresponding reconstructed HSI from the net.

\section{ Simulation Results}
\subsection{Datasets and Metrics}
Three public HSI datasets are used to evaluate GISCnet; they are the ICVL dataset \cite{arad2016sparse}, the CAVE dataset \cite{yasuma2010generalized}, and the Minho dataset \cite{nascimento2002statistics}. The ICVL dataset consists of $201$ HSIs ($1024 \times 1392 \times 31$) and its spectral band is ranged from $400$ $nm$ to $700$ $nm$ with $10$ $nm$ intervals. The CAVE dataset consists of $32$ HSIs with $512  \times 512 \times 31$ pixels; its spectral band is the same as the ICVL dataset. The Minho dataset consists of $30$ HSIs ($820 \times 820 \times 31$), its spectral range  is from $410$ $nm$ to $720$ $nm$ with $10$ $nm$ intervals. In these datasets, we select $15$ channels with spectra ranging from $560$ $nm$ to $700$ $nm$ as our dataset.
  \begin{figure}[htpb]
\centering
\includegraphics[width=6cm,angle=0]{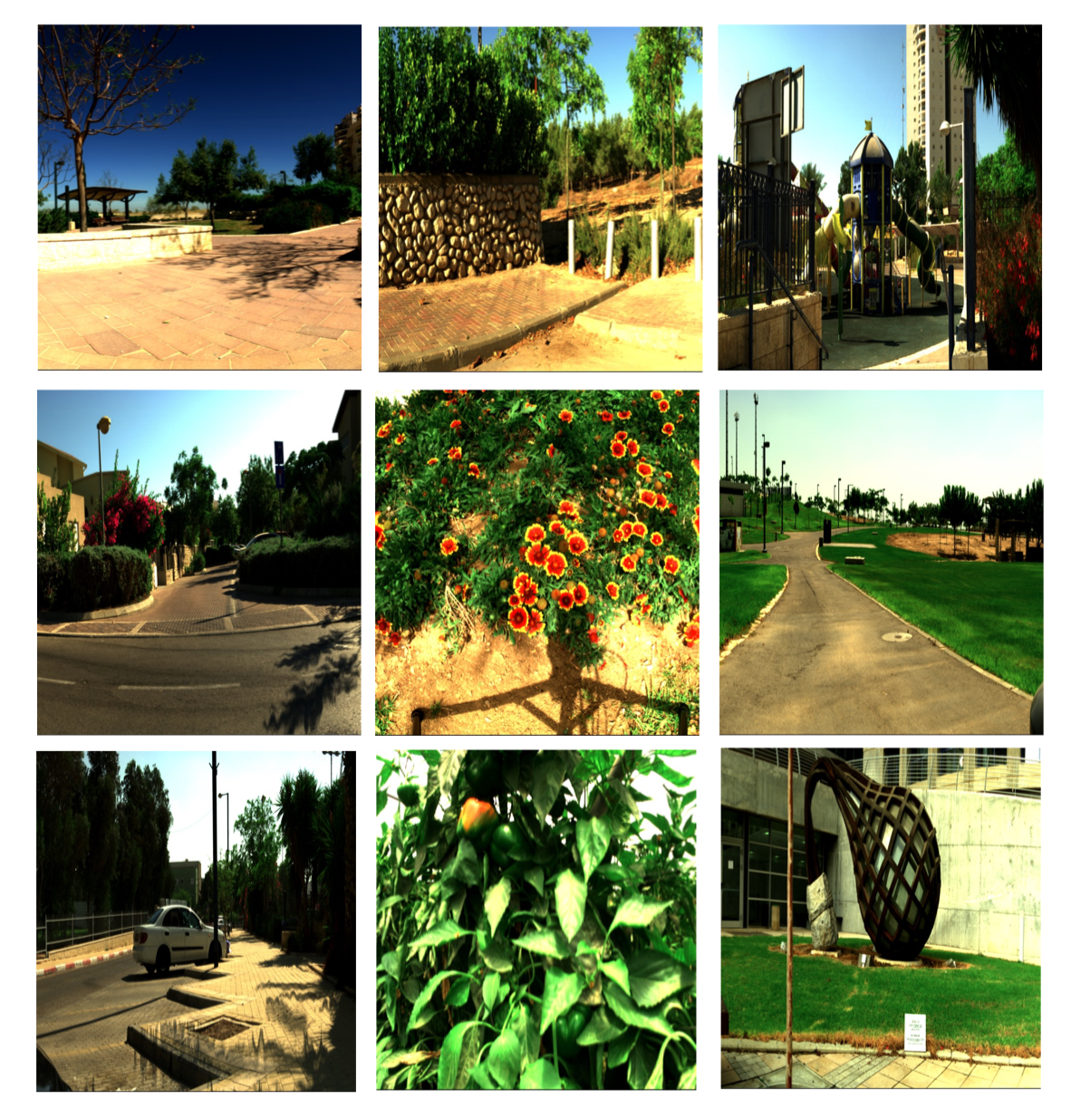}
\caption{The testing figures used in the ICVL dataset\cite{arad2016sparse}.}
\label{ICVL}
\end{figure}



\indent We manually exclude $81$ HSIs with similar backgrounds or contents in the ICVL dataset to eliminate the overfitting effect, then we randomly select $111$ HSIs in the remaining ICVL dataset to generate training and validation sets and thus use the rest of $9$ HSIs for testing. To generate the training and validation sets, firstly, we cut all those $111$ HSIs into slices with size  $144 \times 144 \times 15$, and the cutting stride is set to $144$.  Secondly, $90\%$ slices are randomly selected for training; thus, the rest of the $10\%$ slices are for validation. The nine test ICVL HSIs have been presented in Fig.\ref{ICVL}; we cut them into pieces through the same operation and obtained 648 testing HSI slices with size $144\times 144\times 15$.

All the models are only trained on the ICVL training set, and CAVE and Minho datasets are used as additional testing sets to verify the generalization ability of different networks. Specifically, we selected 27 HSIs in the CAVE dataset and 8 HSIs in the Minho dataset as the second and third testing set, respectively. Using the same cutting operation used in the ICVL dataset, we cut those two testing sets into $243$ CAVE HSI slices and $150$ Minho HSI slices, respectively. PSNR\cite{PSNR}, structural similarity (SSIM)\cite{SSIM}, and spectral angle mapping (SAM) \cite{kruse1993spectral}  are used to evaluate the performance of all methods.


\subsection{Comparisions of Rayleigh and super-Rayleigh speckles}

  \begin{figure}[htpb]
\centering
\includegraphics[width=13cm,angle=0]{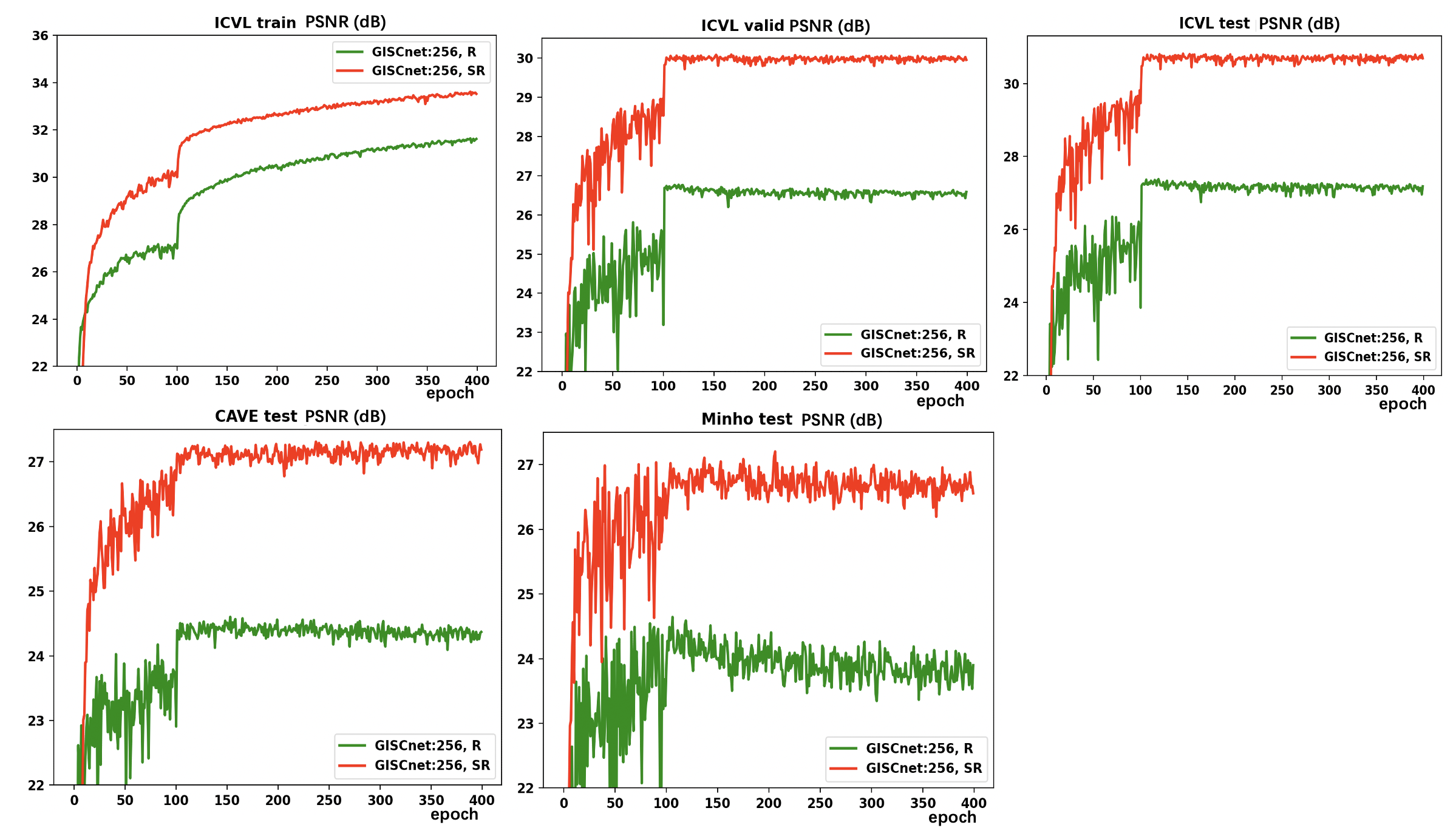}
\caption{The comparison of GISCnet:256 reconstructed curves based on Rayleigh and super-Rayleigh speckles. The ordinate of each subgraph is per (dB), while the abscissa is epoch.}
\label{two}
\end{figure}  

  \begin{figure}[htpb]
\centering
\includegraphics[width=13cm,angle=0]{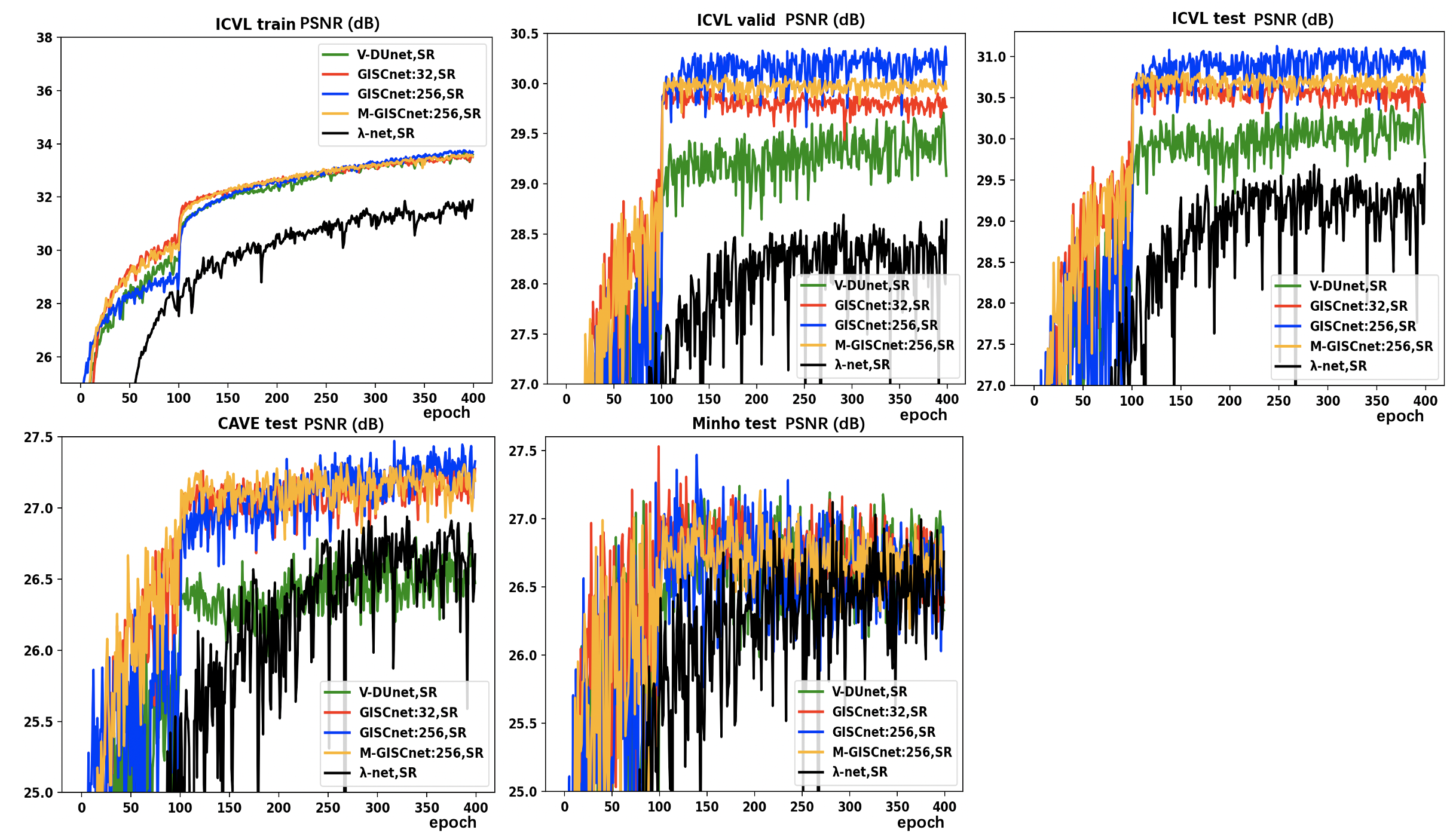}
\caption{The reconstruction results of $\lambda$-net, V-DUnet,  GISCnet:32,M-GISCnet:256 and GISCnet:256  based on super-Rayleigh speckles. The ordinate of each subgraph is per (dB), while the abscissa is epoch. }
\label{five}
\end{figure} 

\begin{table*}
\caption{The performance comparisons on the ICVL, CAVE, and Minho datasets. 648 ICVL HSIs, 243 CAVE HSIs, and 150 Minho HSIs are used to evaluate PSNR, SSIM, and SAM, respectively.}
\label{table_1}
\renewcommand{\arraystretch}{1.8}
\centering
\footnotesize
\begin{tabular}{|c|r|r|r|r|r|r|r|r|r|r|} 
	\hline
	\multirow{2}{*}{Comparison}   &\multicolumn{3}{c|}{ICVL(648)} &\multicolumn{3}{c|}{CAVE(243)} &\multicolumn{3}{c|}{Minho(150)}	 \\ \cline{2-10}  	           
	                 &  PSNR/dB  &   SSIM & SAM   &  PSNR/dB  &   SSIM & SAM &  PSNR/dB  &   SSIM & SAM       \\  \hline     
	    	\multirow{1}{*}{R CS}     
					             &21.6139   & 0.6139 & 0.2372 & 22.0052    & 0.5966 & 0.2977 &20.5951    &  0.5957 & 0.2841 \\   \cline{1-10} 
					             
                \multirow{1}{*}{SR CS}    		                                       &  20.7406    & 0.5637 & 0.2764 & 22.7627    &  0.6105 & 0.2825  & 21.5668    &  0.5666 & 0.3118 \\   \cline{1-10}
                
                \multirow{1}{*}{R GISCnet:256}     
					             & \bf27.6473    &  \bf0.8100 & \bf0.1365 & \bf24.1792    &  \bf0.7497 & \bf0.2447  & \bf24.3802    &  \bf0.7477 & \bf0.2006 \\  \cline{1-10} 	 		
					             			             
		  \multirow{1}{*}{SR GISCnet:256}    		                                       & \bf31.1172    &\bf 0.8946 &\bf0.0932 & \bf27.1723    &  \bf0.8233 & \bf0.1737  &\bf 26.9989    & \bf 0.8262 & \bf0.1559\\   

					             
             \hline    		                                   
\end{tabular}
\end{table*}

\begin{table*}
\caption{The  parameters of GISCnet:32, M-GISCnet:256, and GISCnet:256.}
\label{table_2}
\renewcommand{\arraystretch}{1.8}
\centering
\footnotesize
\begin{tabular}{|c|c|c|c|} 
	\hline
	\multirow{2}{*}{Net}   &\multicolumn{3}{c|}{Parameters} 	 \\ \cline{2-4}   	           
	                   &GR &   UL & TP        \\  \hline     
					             
                \multirow{1}{*}{GISCnet:32}     &  [32,32,32,32,32,32,32,32,32]     & [{\bf32},64,128,256,512,256,128,64,{\bf32}]  & 11.0632M  \\ \cline{1-4}
                
                 \multirow{1}{*}{M-GISCnet:256}      &  [32,32,32,32,32,32,32,32,32]    &[{\bf256},64,128,256,512,256,128,64,{\bf32}]  & 12.2862M \\ \cline{1-4} 
                                                  \multirow{1}{*}{GISCnet:256}      & [32,32,32,32,32,32,32,32,32]     &  [{\bf256},64,128,256,512,256,128,64,{\bf256}] & 14.7432M \\

					             
             \hline    		                                   
\end{tabular}
\end{table*}

\begin{table*}
\caption{The performance comparisons on the ICVL, CAVE, and Minho testing sets. 648 ICVL HSIs, 243 CAVE HSIs, and 150 Minho HSIs are used to evaluate PSNR, SSIM, and SAM, respectively.}
\label{table_3}
\renewcommand{\arraystretch}{1.8}
\centering
\footnotesize
\begin{tabular}{|c|r|r|r|r|r|r|r|r|r|r|} 
	\hline
	\multirow{2}{*}{Comparison}   &\multicolumn{3}{c|}{ICVL(648)} &\multicolumn{3}{c|}{CAVE(243)} &\multicolumn{3}{c|}{Minho(150)}	 \\ \cline{2-10}  	           
	                 &  PSNR/dB  &   SSIM & SAM   &  PSNR/dB  &   SSIM & SAM &  PSNR/dB  &   SSIM & SAM       \\  \hline     
					             
          		
 		 \multirow{1}{*}{SR $\lambda$-net}     
 					             & 29.2877   &  0.8740 & 0.1078 & 26.8025    &  0.7813 & 0.1961  & 26.5672    &  0.8132 & 0.1645 \\  \cline{1-10} 
					             	  
	       \multirow{1}{*}{SR V-DUNet}     
					             & 29.8912    &  0.8843 & 0.1000 & 26.4303    &  0.8069 & 0.1859  & 26.4161    &  0.8173 & 0.1572 \\  \cline{1-10} 	
		 \multirow{1}{*}{SR GISCnet:32}     
					             & 30.7114    &  0.8906 & 0.0997 & 27.0475    &  0.8284 & 0.1854 &27.0325    &  0.8264 & 0.1610 \\  \cline{1-10}
					             
		 \multirow{1}{*}{SR M-GISCnet:256}     
					             & 30.7772   &  0.8923 & 0.0967 & 27.0275    &  0.8292 & 0.1750 & 26.8672    &  0.8245 & 0.1574 \\  \cline{1-10} 				             			                 

					             
					             
		  \multirow{1}{*}{SR GISCnet:256}    		                                       & \bf31.1172    &\bf 0.8946 &\bf0.0932 & \bf27.1723    &  \bf0.8233 & \bf0.1737  &\bf 26.9989    & \bf 0.8262 & \bf0.1559\\   

					             				             
					             
             \hline    		                                   
\end{tabular}
\end{table*}  

\begin{table*}[htpb]
\caption{21 representative reconstructed results on the  ICVL testing set.}
\label{table4}
\renewcommand{\arraystretch}{1.4}
\centering
\footnotesize
\setlength{\tabcolsep}{1mm}{
\begin{tabular}{|cc|c|c|c|c|c|c|c|c|c|c|}
	\hline
	\multicolumn{2}{|c|}{\multirow{2}*{Scene}}  &\multirow{1}*{R  }  &\multirow{1}*{R  }&\multirow{1}*{SR } &\multirow{1}*{SR  } & \multirow{1}*{ SR  } &\multirow{1}*{SR  } &\multirow{1}*{SR } &\multirow{1}*{SR  }  \\ 

&&\multirow{1}*{CS }  &\multirow{1}*{ GISCnet:256 }&\multirow{1}*{CS} &\multirow{1}*{ $\lambda$-net } & \multirow{1}*{ V-DUnet } &\multirow{1}*{GISCnet:32 } &\multirow{1}*{ M-GISCnet:256}&\multirow{1}*{ GISCnet:256  } 	\\  \cline{1-10}   
  
 	\multirow{3}{*}{  \begin{minipage}[b]{0.075\columnwidth}
		\raisebox{-.075\height}{\includegraphics[width=\linewidth]{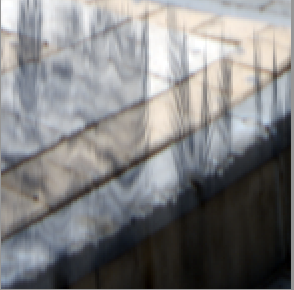}} 
	\end{minipage}
 }                                       &   PSNR/dB        & \bf22.8286 &  \bf28.0378 & \bf18.3271 &  \bf29.0232 & \bf29.1554 &  \bf29.7935 & \bf29.9743 &  \bf30.6608   \\  \cline{2-10}       
		                           &    SSIM        & 0.6851 &  0.8392 & 0.6116 &  0.9080 & 0.9043 &  0.9152 & 0.9101 &  0.9232    \\  \cline{2-10}
					      &   SAM         & 0.1795 &  0.1001 & 0.2593 &  0.0783 & 0.0794 &  0.0821 & 0.0798 &  0.0741     \\\cline{1-10} 
					            					            
 	\multirow{3}{*}{\begin{minipage}[b]{0.075\columnwidth}
		\raisebox{-.075\height}{\includegraphics[width=\linewidth]{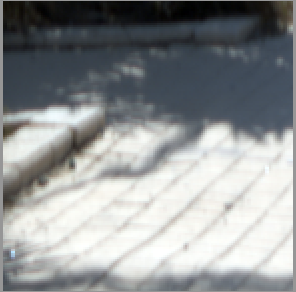}} 
	\end{minipage}}            &   PSNR/dB        & \bf23.5773    &  \bf27.3689    & \bf17.5298  & \bf27.1092   & \bf30.5706   & \bf32.0377  & \bf32.0613    & \bf32.5753    \\  \cline{2-10}       
		                           &    SSIM        & 0.6508    &  0.8464   & 0.6606   & 0.9180   &0.9229   &0.9334  & 0.9346    & 0.9377    \\  \cline{2-10}
					      &   SAM         & 0.1345    &  0.0751   & 0.1503    &0.0534    & 0.0557   &0.0483  & 0.0481    &0.0460      \\\cline{1-10} 
	 	
		\multirow{3}{*}{\begin{minipage}[b]{0.075\columnwidth}
		\raisebox{-.075\height}{\includegraphics[width=\linewidth]{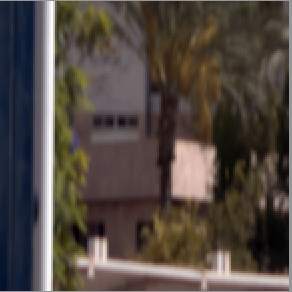}} 
	\end{minipage}}            &   PSNR/dB        & \bf23.4632 &  \bf28.3170 & \bf23.0792 &  \bf29.9017 & \bf30.0373 &  \bf30.4557 & \bf30.8289 &  \bf31.2644     \\  \cline{2-10}       
		                           &    SSIM        & 0.6544 &  0.7843 & 0.6572 &  0.8764 & 0.8779 &  0.8874 & 0.8886 &  0.8883    \\  \cline{2-10}
					      &   SAM         & 0.2525 &  0.1795 & 0.2871 &  0.1317 & 0.1315 &  0.1308 & 0.1257 &  0.1298       \\\cline{1-10} 
		 	
		\multirow{3}{*}{\begin{minipage}[b]{0.075\columnwidth}
		\raisebox{-.075\height}{\includegraphics[width=\linewidth]{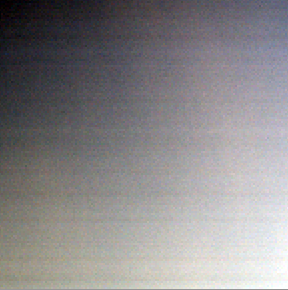}} 
	\end{minipage}}            &   PSNR/dB        & \bf12.7716 &  \bf32.4965 & \bf12.4161 &  \bf28.1162 & \bf32.5826 &  \bf33.1589 & \bf33.2044 & \bf 34.2210   \\  \cline{2-10}       
		                           &    SSIM        & 0.3769 &  0.8291 & 0.2339 &  0.8204 & 0.8346 &  0.8209 & 0.8290 &  0.8344    \\  \cline{2-10}
					      &   SAM         & 0.2845 &  0.0395 & 0.3176 &  0.0537 & 0.0357 &  0.0405 & 0.0377 &  0.0348       \\\cline{1-10} 
					      
		 	\multirow{3}{*}{\begin{minipage}[b]{0.075\columnwidth}
		\raisebox{-.075\height}{\includegraphics[width=\linewidth]{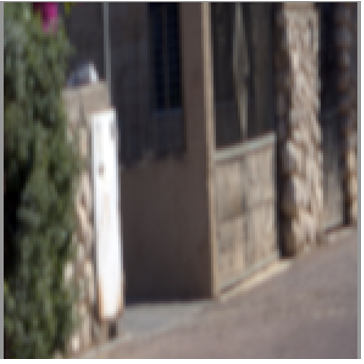}} 
	\end{minipage}}            &   PSNR/dB        &\bf 22.5280 &  \bf27.5408 & \bf23.5837 &  \bf30.4246 & \bf31.4579 &  \bf31.7284 & \bf32.1504 &  \bf32.7506     \\  \cline{2-10}       
		                           &    SSIM        & 0.6309 &  0.8018 & 0.6937 &  0.9040 & 0.9056 &  0.9083 & 0.9140 &  0.9169    \\  \cline{2-10}
					      &   SAM         & 0.2799 &  0.1603 & 0.2221 &  0.1023 & 0.0960 &  0.0964 & 0.0904 &  0.0868      \\\cline{1-10} 
					            					            
 	\multirow{3}{*}{\begin{minipage}[b]{0.075\columnwidth}
		\raisebox{-.075\height}{\includegraphics[width=\linewidth]{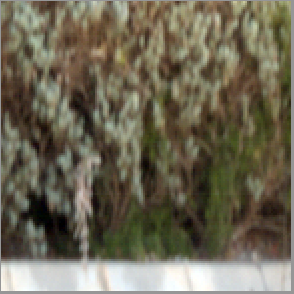}} 
	\end{minipage}}            &   PSNR/dB        & \bf23.4131 &  \bf26.5331 & \bf19.0805 &  \bf27.5319 & \bf28.6497 &  \bf29.3085 & \bf29.3589 & \bf29.4968     \\  \cline{2-10}       
		                           &    SSIM        & 0.6473 &  0.7392 & 0.5908 &  0.8515 & 0.8561 &  0.8583 & 0.8608 &  0.8653    \\  \cline{2-10}
					      &   SAM         & 0.2150 &  0.1591 & 0.3060 &  0.1335 & 0.1114 &  0.1167 & 0.1218 &  0.1045      \\\cline{1-10} 
	 	
		\multirow{3}{*}{\begin{minipage}[b]{0.075\columnwidth}
		\raisebox{-.075\height}{\includegraphics[width=\linewidth]{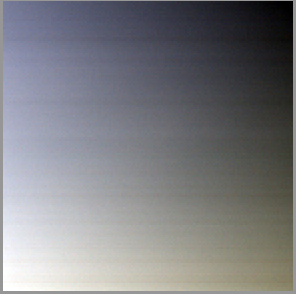}} 
	\end{minipage}}            &   PSNR/dB        & \bf14.2832    &  \bf34.1381   & \bf14.1206   & \bf29.8432   & \bf30.9348   &\bf 31.9191  &\bf 34.3543    & \bf35.7079    \\  \cline{2-10}       
		                           &    SSIM        & 0.3844   &  0.9328   & 0.2503    & 0.9214   & 0.9385   & 0.9227 & 0.9366    &0.9382    \\  \cline{2-10}
					      &   SAM         &0.2850    &  0.0364   & 0.3054    & 0.0465  & 0.0326   & 0.0401 & 0.0353    & 0.0299      \\\cline{1-10} 
		 	
		\multirow{3}{*}{\begin{minipage}[b]{0.075\columnwidth}
		\raisebox{-.075\height}{\includegraphics[width=\linewidth]{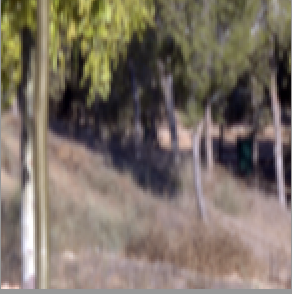}} 
	\end{minipage}}            &   PSNR/dB        & \bf21.6152    & \bf 25.0434    & \bf21.0931   & \bf28.5385   & \bf29.3571  &\bf 29.5028  & \bf29.5687   & \bf29.5693    \\  \cline{2-10}       
		                           &    SSIM        & 0.6480    &  0.7514   & 0.6635    & 0.8873   & 0.8930   & 0.8960  & 0.8907    &0.8995    \\  \cline{2-10}
					      &   SAM         & 0.2496    &  0.1733  &0.2738   & 0.1207    & 0.1060   & 0.1123 & 0.1096    & 0.1080      \\\cline{1-10} 
					      
		 	\multirow{3}{*}{\begin{minipage}[b]{0.075\columnwidth}
		\raisebox{-.075\height}{\includegraphics[width=\linewidth]{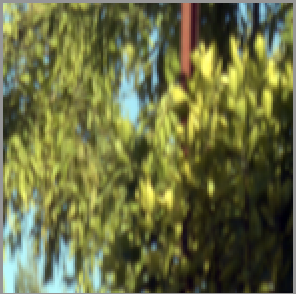}} 
	\end{minipage}}            &   PSNR/dB        &\bf 22.9986 &  \bf26.6498 & \bf25.3873 & \bf 30.0282 & \bf30.3802 &  \bf30.3907 & \bf31.2143 & \bf 31.2600    \\  \cline{2-10}       
		                           &    SSIM        & 0.6587 &  0.7854 & 0.7811 &  0.9131 & 0.9118 &  0.9180 & 0.9229 &  0.9240     \\  \cline{2-10}
					      &   SAM         & 0.3236    & 0.2221   & 0.2621  & 0.1386    &0.1424   & 0.1396  & 0.1307    & 0.1359      \\\cline{1-10} 
					            					            
 	\multirow{3}{*}{\begin{minipage}[b]{0.075\columnwidth}
		\raisebox{-.075\height}{\includegraphics[width=\linewidth]{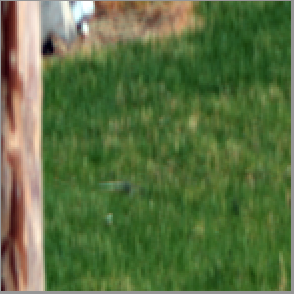}} 
	\end{minipage}}            &   PSNR/dB        & \bf22.3796 &  \bf25.8715 & \bf21.9656 &  \bf27.6980 & \bf28.5794 &  \bf28.8485 & \bf29.7029 &  \bf29.7800   \\  \cline{2-10}       
		                           &    SSIM        & 0.6258 &  0.7330 & 0.5747 &  0.8506 & 0.8551 &  0.8517 & 0.8500 &  0.8600     \\  \cline{2-10}
					      &   SAM         & 0.2776 &  0.2050 & 0.2801 &  0.1458 & 0.1239 &  0.1372 & 0.1326 &  0.1307      \\\cline{1-10} 
	 	
		\multirow{3}{*}{\begin{minipage}[b]{0.075\columnwidth}
		\raisebox{-.075\height}{\includegraphics[width=\linewidth]{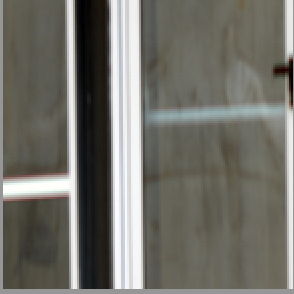}} 
	\end{minipage}}            &   PSNR/dB        & \bf21.8101 &  \bf26.2900 & \bf18.1914 &  \bf27.3651 & \bf27.7308 &  \bf28.5930 & \bf28.6141 &  \bf29.7817   \\  \cline{2-10}       
		                           &    SSIM        & 0.5667 &  0.8043 & 0.5015 &  0.8719 & 0.8978 &  0.8885 & 0.8972 &  0.8997    \\  \cline{2-10}
					      &   SAM         & 0.2391 &  0.1427 & 0.2901 &  0.1274 & 0.1119 &  0.1099 & 0.1142 &  0.0990     \\\cline{1-10} 
					      
    \multirow{3}{*}{\begin{minipage}[b]{0.075\columnwidth}
		\raisebox{-.075\height}{\includegraphics[width=\linewidth]{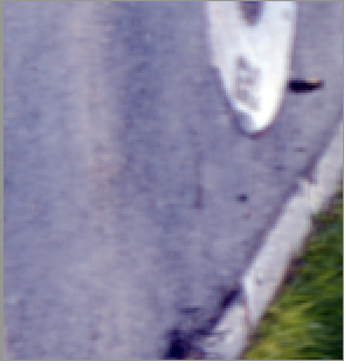}} 
	\end{minipage}}            &   PSNR/dB        & \bf21.1965 &  \bf28.1659 & \bf15.4773 &  \bf27.0289 & \bf30.3679 &  \bf30.5159 & \bf31.4309 &  \bf31.8214   \\  \cline{2-10}       
		                           &    SSIM      & 0.5348 &  0.8241 & 0.3441 &  0.8611 & 0.8715 &  0.8657 & 0.8742 &  0.8755   \\  \cline{2-10}
					      &   SAM         & 0.2065 &  0.0929 & 0.3487 &  0.0826 & 0.0636 &  0.0721 & 0.0637 &  0.0660      \\\cline{1-10}
					      
	    \multirow{3}{*}{\begin{minipage}[b]{0.075\columnwidth}
		\raisebox{-.075\height}{\includegraphics[width=\linewidth]{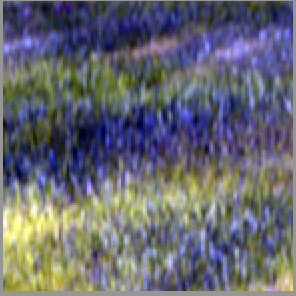}} 
	\end{minipage}}            &   PSNR/dB        & \bf20.1245 &  \bf25.9931 & \bf20.3473 &  \bf28.4795 & \bf30.1561 &  \bf30.3651 & \bf30.3890 &  \bf31.1648   \\  \cline{2-10}       
		                           &    SSIM        & 0.5691 &  0.6859 & 0.5411 &  0.8291 & 0.8347 &  0.8351 & 0.8455 &  0.8538     \\  \cline{2-10}
					      &   SAM         & 0.2618 &  0.1337 & 0.2594 &  0.1230 & 0.0970 &  0.1015 & 0.1069 &  0.0859      \\\cline{1-10}
					      
		    \multirow{3}{*}{\begin{minipage}[b]{0.075\columnwidth}
		\raisebox{-.075\height}{\includegraphics[width=\linewidth]{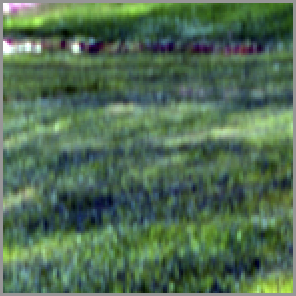}} 
	\end{minipage}}            &   PSNR/dB        & \bf21.8021 &  \bf28.7269 & \bf20.9752 &  \bf28.0102 & \bf30.2205 &  \bf30.7422 & \bf31.2361 &  \bf31.8664    \\  \cline{2-10}       
		                           &    SSIM        & 0.6171 &  0.7787 & 0.4967 &  0.8426 & 0.8497 &  0.8438 & 0.8550 &  0.8643   \\  \cline{2-10}
					      &   SAM         & 0.2592 &  0.1429 & 0.2926 &  0.1488 & 0.1120 &  0.1281 & 0.1247 &  0.1061      \\\cline{1-10} 
 	\multirow{3}{*}{\begin{minipage}[b]{0.075\columnwidth}
		\raisebox{-.075\height}{\includegraphics[width=\linewidth]{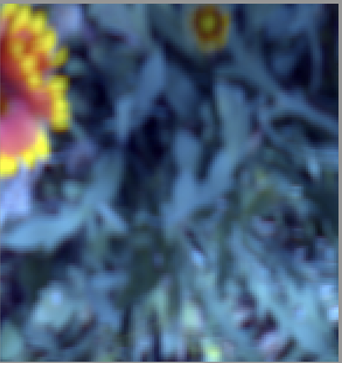}} 
	\end{minipage}}            &   PSNR/dB        & \bf22.4698 & \bf 26.3207 & \bf27.4845 &  \bf30.9256 & \bf32.3063 &  \bf32.4058 &\bf 33.0834 & \bf 33.9067     \\  \cline{2-10}       
		                           &    SSIM        & 0.6346 &  0.7979 & 0.7035 &  0.9033 & 0.8876 &  0.9014 & 0.9073 &  0.9132    \\  \cline{2-10}
					      &   SAM         &0.3386 &  0.2323 & 0.2781 &  0.1459 & 0.1495 &  0.1463 & 0.1412 &  0.1370      \\ \cline{1-10} 
					      
\multicolumn{10}{|l|}{\small\sl continued on next page}\\					      
\hline

\end{tabular}}
\end{table*}

 \begin{table*}[htpb]
\renewcommand{\arraystretch}{1.4}
\centering
\footnotesize
\setlength{\tabcolsep}{1mm}{
\begin{tabular}{|cc|c|c|c|c|c|c|c|c|c|c|}
	\hline
       \multicolumn{10}{|l|}{\small\sl continued from previous page}\\	  \cline{1-10}       	
	\multicolumn{2}{|c|}{\multirow{2}*{Scene}}  &\multirow{1}*{R  }  &\multirow{1}*{R  }&\multirow{1}*{SR } &\multirow{1}*{SR  } & \multirow{1}*{ SR  } &\multirow{1}*{SR  } &\multirow{1}*{SR } &\multirow{1}*{SR  }  \\ 

&&\multirow{1}*{CS }  &\multirow{1}*{ GISCnet:256 }&\multirow{1}*{CS} &\multirow{1}*{ $\lambda$-net } & \multirow{1}*{ V-DUnet } &\multirow{1}*{GISCnet:32 } &\multirow{1}*{ M-GISCnet:256} &\multirow{1}*{ GISCnet:256  } 	\\  \cline{1-10}   
  
 	\multirow{3}{*}{\begin{minipage}[b]{0.075\columnwidth}
		\raisebox{-.075\height}{\includegraphics[width=\linewidth]{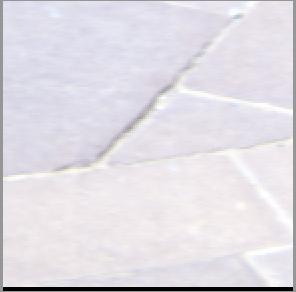}} 
	\end{minipage}}            &   PSNR/dB        & \bf14.0936 &  \bf30.2179 & \bf8.7763 & \bf 27.8679 &\bf 30.0199 &  \bf31.4096 & \bf31.8608 &  \bf31.8659   \\  \cline{2-10}       
		                           &    SSIM        & 0.3510 &  0.7413 & 0.2357 &  0.8116 & 0.8169 &  0.8197 & 0.8242 &  0.8306    \\  \cline{2-10}
					      &   SAM         & 0.2076 &  0.0452 & 0.3450 &  0.0468 & 0.0384 &  0.0367 & 0.0369 &  0.0341      \\\cline{1-10} 
	 	
		\multirow{3}{*}{\begin{minipage}[b]{0.075\columnwidth}
		\raisebox{-.075\height}{\includegraphics[width=\linewidth]{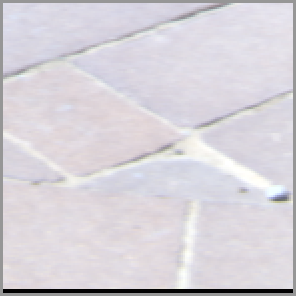}} 
	\end{minipage}}            &   PSNR/dB        &\bf16.5169 &  \bf29.2259 & \bf10.1530 &  \bf29.8301 & \bf30.1389 &  \bf31.6060 & \bf32.0056 &  \bf32.4432   \\  \cline{2-10}       
		                           &    SSIM        & 0.3881 &  0.7601 & 0.2609 &  0.8281 & 0.8391 &  0.8361 & 0.8454 &  0.8461    \\  \cline{2-10}
					      &   SAM         & 0.2126 &  0.0530 & 0.3477 &  0.0480 & 0.0418 &  0.0408 & 0.0391 &  0.0377      \\\cline{1-10} 
					      					      
		\multirow{3}{*}{\begin{minipage}[b]{0.075\columnwidth}
		\raisebox{-.075\height}{\includegraphics[width=\linewidth]{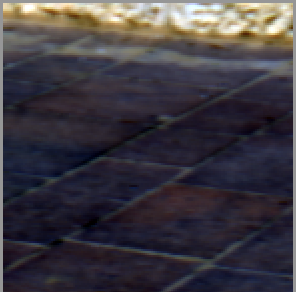}} 
	\end{minipage}}            &   PSNR/dB        & \bf22.9186 &  \bf30.0261 & \bf21.8255 &  \bf30.9852 & \bf32.5824 &  \bf33.7393 & \bf34.2162 &  \bf34.5561     \\  \cline{2-10}       
		                           &    SSIM        & 0.5877 &  0.8514 & 0.4917 &  0.9062 & 0.9087 &  0.9096 & 0.9143 &  0.9157   \\  \cline{2-10}
					      &   SAM         & 0.2749 &  0.1688 & 0.2970 &  0.1381 & 0.1078 &  0.1608 & 0.1257 &  0.1080      \\\cline{1-10} 
					      
		 	\multirow{3}{*}{\begin{minipage}[b]{0.075\columnwidth}
		\raisebox{-.075\height}{\includegraphics[width=\linewidth]{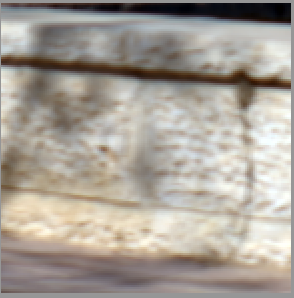}} 
	\end{minipage}}            &   PSNR/dB        & \bf19.3742 &  \bf24.9192 & \bf13.5305 &  \bf27.1389 & \bf29.0044 &  \bf29.4488 & \bf29.8613 & \bf 30.0706    \\  \cline{2-10}       
		                           &    SSIM        & 0.6217 &  0.7736 & 0.4760 &  0.8876 & 0.8893 &  0.8908 & 0.8913 &  0.8944    \\  \cline{2-10}
					      &   SAM         & 0.2004 &  0.0919 & 0.2966 &  0.0658 & 0.0640 &  0.0619 & 0.0633 &  0.0610      \\\cline{1-10} 
					            					            
 	\multirow{3}{*}{\begin{minipage}[b]{0.075\columnwidth}
		\raisebox{-.075\height}{\includegraphics[width=\linewidth]{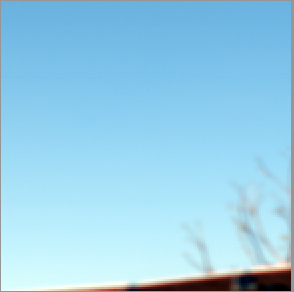}} 
	\end{minipage}}            &   PSNR/dB        & \bf18.1265 &  \bf27.5425 & \bf14.2699 &  \bf27.9522 & \bf30.2986 &  \bf31.9711 & \bf32.6159 &  \bf32.6808     \\  \cline{2-10}       
		                           &    SSIM        & 0.4402 &  0.9196 & 0.2787 &  0.8985 & 0.9292 &  0.9208 & 0.9362 &  0.9373    \\  \cline{2-10}
					      &   SAM         & 0.2170 &  0.0601 & 0.3032 &  0.0687 & 0.0582 &  0.0527 & 0.0482 &  0.0486       \\\cline{1-10} 
	 	
		\multirow{3}{*}{\begin{minipage}[b]{0.075\columnwidth}
		\raisebox{-.075\height}{\includegraphics[width=\linewidth]{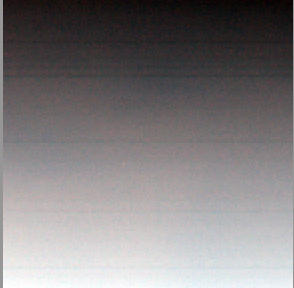}} 
	\end{minipage}}            &   PSNR/dB        & \bf16.0620 &  \bf34.1044 & \bf15.7644 & \bf 30.4821 & \bf34.7885 &  \bf35.2209 & \bf35.7304 &  \bf35.8687    \\  \cline{2-10}       
		                           &    SSIM        & 0.4335 &  0.8918 & 0.2882 &  0.8813 & 0.8925 &  0.8788 & 0.8898 &  0.8919    \\  \cline{2-10}
					      &   SAM         & 0.2643 &  0.0320 & 0.2993 &  0.0476 & 0.0329 &  0.0435 & 0.0380 &  0.0309      \\\cline{1-10}

\end{tabular}}
\end{table*}

Firstly, we apply two calibration speckles in the GISC spectral camera to see their influence on the network reconstruction results: Rayleigh speckles and super-Rayleigh speckles. Their comparison results have been displayed in Fig.\ref{two} and Table \ref{table_1}. It should be noted that SR is super-Rayleigh's abbreviation, and R is Rayleigh's abbreviation in all figures and tables. From Fig.\ref{two} and Table \ref{table_1}, we can see that the reconstructed results based on super-Rayleigh speckles are far better than that based on Rayleigh speckles.  It can be seen from Table \ref{table_1}  that although the evaluation values of the average CS preprocessing image based on super-Rayleigh speckles and Rayleigh speckles are nearly the same,  strictly speaking, the average evaluations based on super-Rayleigh speckles are even worse than those found on Rayleigh speckles. Still, the network reconstruction results based on the super-Rayleigh speckles are much better than those based on the Rayleigh speckles. After the average evaluations, we get its average evaluation value of PSNR/SSIM/SAM is 31.11 dB/0.89/0.09 (using super-Rayleigh speckles) and 27.64 dB /0.74 /0.24 (using Rayleigh speckles) on 648 ICVL testing HSIs, 27.17 dB /0.82 /0.17 (using super-Rayleigh speckles) and 24.17 dB /0.82 /0.17 (using Rayleigh speckles) on 243 CAVE testing HSIs, and 26.99 dB/ 0.82 /0.15 (using super-Rayleigh speckles) and 24.38 dB /0.74 /0.20 (using Rayleigh speckles) on 150 Minho testing HSIs. So compared with the case of using Rayleigh speckles in the GISC spectral camera,  the average PSNR with super-Rayleigh speckles has been improved by about 3.5 dB on the ICVL testing set, about 3 dB on the CAVE testing set, and about 2.6 dB on the Minho testing set. Therefore, the application of super-Rayleigh speckles can significantly improve the reconstructed image quality of the GISC spectral camera. The reason the average evaluation values of reconstructed results on the ICVL testing set are far better than that on the other two testing sets is also evident because the HSIs used in the training set are only selected from the ICVL dataset.
In contrast, the HSI data of the other two datasets are entirely excluded from the training set. And this is a common failing of almost all data-driven networks in that their generalization ability is limited. Despite all this, there is still a gap between the network generalization ability of different networks.

\subsection{Comparisions of different networks}

       Secondly, to verify the performance of our proposed GISCnet and to see its generalization ability, we compare it with two representative reconstruction networks, including V-DUNet and $\lambda$-net \cite{miao2019net}. V-DUnet is specially designed for GISC spectral camera while $\lambda$-net is designed for CASSI \cite{arce2013compressive,yuan2021snapshot}. Although $ \lambda $- net is not intended for GISC spectral camera, we still try to apply it to GISC spectral camera.  $\lambda$-net  also contains two inputs: the mask and detection image. We need to replace the mask image with the preprocessed CS image, then $\lambda$-net can be directly applied to GISC spectral camera.  For GISCnet, we compare its reconstruction results under different parameter settings (mainly for parameters of the Unet layer). As we all know, the changing rule of the Unet layer is to gradually decrease by a multiple of 2 and then increase by a multiple of 2,  [32, 64, 128, 256, 512, 256, 128, 64, 32] is a typical conventional Unet layer parameter. In the GISCnet, we additionally test GISCnet under two unconventional Unet layer parameters; see Table \ref{table_2} for specific parameter settings of GISCnet.

 The GR in Table \ref{table_2} denotes the growth rate parameter in the Dense block, UL is the Unet layer parameter, and TP represents the training parameter of GISCnet. We denote GISCnet with regular Unet layer parameters by GISCnet:32, the GISCnet with Unet layer parameters [256, 64, 128, 256, 512, 256, 128, 64, 32] is represented by M-GISCnet:256, while the GISCnet with Unet layer parameters [256, 64, 128, 256, 512, 256, 128, 64, 256] is represented by GISCnet:256.  And the comparison results between V-DUNet, $\lambda$-net, and GISCnet are shown in Fig.\ref{five} and Table \ref{table_3}.  
 As can be seen in  Fig.\ref{five} and Table \ref{table_3}, GISCnet (whether GISCnet:32, M-GISCnet:256 or GISCnet:256) not only has better reconstruction results than V-DUnet and $\lambda$-net but also has higher generalization ability than them (with better reconstruction performance on the Minho and CAVE testing sets). So this has validated that: although the generalization ability of model-driven networks is limited, there are still differences between different network models. And in GISCnet, it is evident that the reconstruction result of GISCnet:256 is better than M-GISCnet:256, which in turn is better than GISCnet:32. Therefore, we can choose GISCnet:256 as our final solution if we want the reconstruction result to be as good as possible.  Perhaps because $\lambda$-net is not designed for GISC spectral cameras, its reconstruction result is worse than those of GISCnet and V-DUnet, designed explicitly for the GISC spectral camera.

     \begin{figure*}[htpp]
\centering
\includegraphics[width=13cm,angle=0]{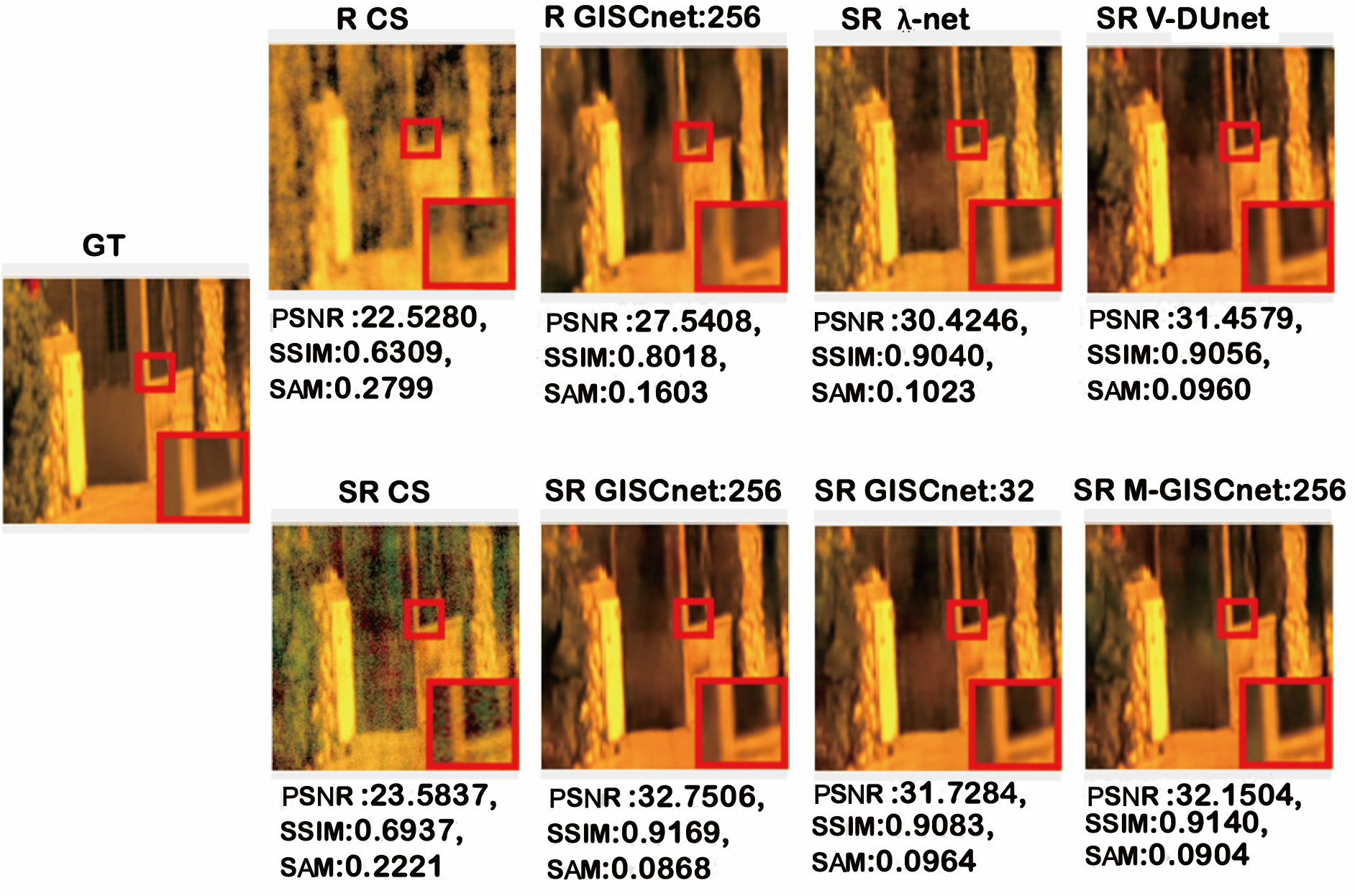}
\caption{Exemplar reconstructed image (selected from Table.\ref{table4}).}
\label{weicai1}
\end{figure*}   

     \begin{figure*}[htpp]
\centering
\includegraphics[width=13cm,angle=0]{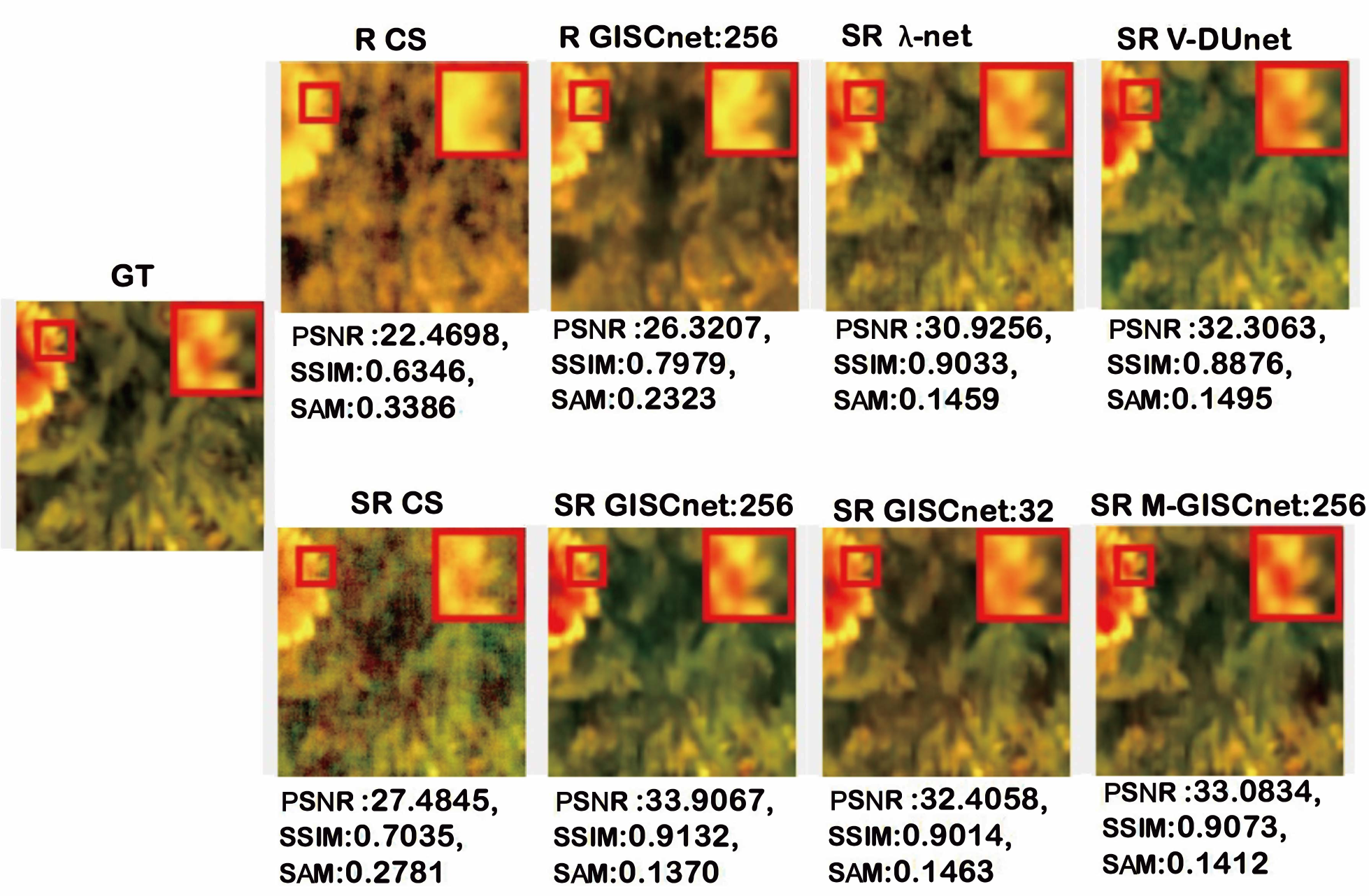}
\caption{Exemplar reconstructed image (selected from Table.\ref{table4}).}
\label{weicai2}
\end{figure*}   
 
Although we have averaged 648 HSIs on the ICVL testing set to evaluate the network performance, due to the article's length, we can't display all those 648 HSIs, so we only select some representative HSIs to show the reconstruction results. From Fig.\ref{five} and Table \ref{table_3}, we can see that the average evaluated PSNR of different networks meets the following rules: $\lambda$-net< V-DUnet < GISCnet:32 < M-GISCnet:256 < GISCnet:256. However, the probability that each image's PSNR on the above networks can satisfy this averaging rule in these 648 testing HSIs is very low. In fact, out of those 648 ICVL testing HSIs, we find only 52 HSIs' PSNRs that meet this averaging rule. If we are to show all 52 HSIs, it will still take up too much space, so we remove those images with too large or too small PSNR values: taking $\lambda$-net as the benchmark, we kept only those HSIs with $\lambda$-net PSNR values in the range of 27$\leq $$\lambda$-net$\leq$31. Then we finally obtained 21 HSIs; their results are shown in Table \ref{table4}. Two images are selected from Table \ref{table4} to visualize the reconstruction results further, and the results are presented in Fig.\ref{weicai1} and Fig.\ref{weicai2}. In Fig.\ref{weicai1} and Fig.\ref{weicai2}, instead of showing all 15 channel images of these HSIs, we have integrated all 15 channel images into a single pseudo-color image for display\cite{weicai}. {{The results of averaging a large number of HSIs are more convincing than those of a small number of HSIs, so for the evaluation of network performance, the results displayed in Fig.\ref{two}, Fig.\ref{five}, Table \ref{table_1} and Table \ref{table_3} are more convincing than Table \ref{table4}, Fig.\ref{weicai1} and Fig.\ref{weicai2}.}}

 Finally, although we only test GISCnet under three different parameters, GISCnet can be set to many other parameters (its growth rate and Unet layer parameters can be adjusted freely). That is, many cases of GISCnet with different parameters can be tested. If one tries more network reconstruction results with additional parameters, one may get better reconstruction results.

\section{Conclusion} 
We proposed an end-to-end GISCnet to improve the image reconstruction quality in GISC spectral camera. Two calibration speckles are applied in the GISC spectral camera to see their influence on the network reconstruction results: Rayleigh speckles and super-Rayleigh speckles. Compared with the case of using Rayleigh speckles in the GISC spectral camera, the average PSNR with super-Rayleigh speckles has dramatically improved by about 3.5 dB on the ICVL testing set, about 3 dB on the CAVE testing set, and about 2.6 dB on the Minho testing set. Compared with other reconstructed networks, GISCnet has better image reconstruction quality and network generalization ability for GISC spectral camera. Since GISCnet is mainly designed by UD blocks, its parameters can be set freely. We demonstrate three different parameters of GISCnet models and compare their reconstruction results. It shows that the GISCnet’s results can be further improved by optimizing its parameters, and we will make more efforts on the parameters’ optimization. In general, this paper shows that optimizing light field modulation and deep learning image reconstruction can effectively improve the imaging quality of the system. Therefore, our subsequent research work is further joint optimization of light-field modulation and image reconstruction through deep learning.

}

{
    \small
    \bibliographystyle{ieeenat_fullname}
    \bibliography{GISCnet}
}


\end{document}